\documentclass[a4paper,envcountsame]{llncs}

\pdfoutput=1

\usepackage{xspace}
\usepackage{latexsym}
\usepackage{amssymb}   

\usepackage{ifthen}
\newboolean{commentsaon}         
\setboolean{commentsaon}{true}
  \setboolean{commentsaon}{false}  

\newboolean{commentson} 
\setboolean{commentson}{true}

\newcommand{\comment}[1]
{\ifthenelse{\boolean{commentson}\AND\boolean{commentsaon}}
   {{\par\noindent\mbox{}{\small\blue[ *** #1 ]\par}\noindent\par}}{}}

\newcommand{\commenta}[1]
{\ifthenelse{\boolean{commentsaon}}
   {{\par\noindent\mbox{}{\small\color[rgb]{0, .5, 0}[ *** #1 ]\par}\noindent\par}}{}}

\usepackage[yyyymmdd]{datetime}

 \newcommand{\myhalfmagnification}
{0.002}
\usepackage{color}
\newcommand\blue     {\color{blue}}

\usepackage{hyperref}
\usepackage[hyphenbreaks]{breakurl}
\usepackage{pgf}

\newtheorem{algorithm}{Algorithm}

\newcommand*{\seq}[2][n]  {{#2_{1}, \allowbreak \ldots, \allowbreak #2_{#1}}}

\newcommand*{\notmodels}{\mathrel{\,\not\!\models}}

\usepackage[mathscr]{euscript}

\newcommand*{\TT}{{\ensuremath{\mathcal{T}}}\xspace}

\newcommand*{\enter}{\fbox{\small enter}\xspace}
\newcommand*{\rr}{\mbox{\bf r}\xspace}
\newcommand*{\cc}{\mbox{\bf c}}

\newcommand*{\jj}{\mbox{\bf j}}
\renewcommand*{\ss}{\mbox{\bf s}\xspace}
\newcommand*{\dmessage}[4]
    {{\ensuremath{#1\quad #2\,\ \mbox{\tt #3:}\ #4}\xspace}}
\newcommand*{\Call}[3]{\dmessage {#1} {#2} {Call} {#3}}
\newcommand*{\Exit}[3]{\dmessage {#1} {#2} {Exit} {#3}}
\newcommand*{\Redo}[3]{\dmessage {#1} {#2} {Redo} {#3}}
\newcommand*{\Fail}[3]{\dmessage {#1} {#2} {Fail} {#3}}

\begin{document}

\title{The Prolog Debugger \\ and Declarative Programming%
       \vspace{-2ex}%
      }
\DeclareRobustCommand{\mytitlecomment}
    {\ifthenelse{\boolean{commentson}\AND\boolean{commentsaon}}
        {\blue\ [v.3.2, \today, with private comments]}{}
    }
\author   {%
           W{\l}odzimierz Drabent\inst{1,2}%
           \\[1ex]
           2020-02-08
           \vspace{-1ex}%
            \mytitlecomment}

\institute{Institute of Computer Science,
         Polish Academy of Sciences
    \email{drabent\,{\it at}\/\,ipipan\,{\it dot}\/\,waw\,{\it dot}\/\,pl}
         \and
         IDA, Link\"oping University, Sweden%
      \vspace{-2ex}%
  }

\maketitle 

\newcommand*{\myvspace}{\vspace{-.7ex}}

\enlargethispage{1.5ex}
\myvspace\myvspace\myvspace\myvspace

\begin{abstract}
  Logic programming is a declarative programming paradigm.  
Programming language Prolog makes logic programming possible, at least to a
substantial extent.
However the Prolog debugger works solely in terms of the
  operational semantics.  So it is incompatible with declarative programming.
  This report discusses this issue and tries to find how the debugger may be
  used from the declarative point of view.
  The results are rather not encouraging.

  Also,
  the box model of Byrd, used by the debugger, is explained in terms of
  SLD-resolution. 
\myvspace
\myvspace\myvspace
\end{abstract}

\section{Introduction}
The idea of logic programming is that a program is a set of logic formulae,
and a computation means producing logical
consequences of the program.  
So it is a {\em declarative} programming paradigm.  The program is not a
description of any computation, it may be rather seen as a description of a
problem to solve.  Answers of a given program (the {\em logic}) may be computed
under various strategies (the {\em control}\/), the results depend solely on
the former.
This semantics of programs, based on logic, is called {\em declarative
semantics}.

Programming language Prolog is a main implementation of logic programming.
Its core, which may be called ``pure Prolog'',
 is an implementation of SLD-resolution under a fixed control.
(SLD-resolution with Prolog selection rule is called LD-resolution.)
For a given program $P$ and query $Q$,
Prolog computes
logical consequences of $P$ which are instances of $Q$.
If the computation is finite then, roughly speaking, all such consequences
are computed.%
\footnote{%
See e.g.\ \cite{Apt-Prolog} for details.
We omit the issue of unification without occur-check; it may lead to
incorrect answers 
(i.e.\ not being logical consequences of the program). 

Technically, by an answer of a program we mean the result of applying a
(correct or computed) answer substitution to a query.
}

On the other hand, Prolog may be viewed 
   without any reference to logic, 
  as a programming language with a specific control flow, the terms as the
  data, and a certain kind of term matching as the main primitive operation. 
  Such a view is even necessary when we deal with non logical features of
  full Prolog, like the built-ins dealing with input/output.
  Of course such {\em operational view} loses all the advantages 
  of declarative programming.

In the author's opinion,
Prolog makes declarative programming possible in practice. 
  A Prolog program treated as a set of logical
clauses is a logic program.  The logic determines the answers of the program.
At a lower level, the programmer can influence the control.  This can be
done by setting the order of program clauses and the order of premises within
a clause (and by some additional Prolog constructs).  Changing the control 
keeps the logic intact, and thus the program's answers are unchanged;
the logic is separated from the control \cite{DBLP:journals/cacm/Kowalski79}. 
What is changed is the way they are computed, for instance the computation
may be made more efficient.  In particular, an infinite computation may be
changed into a finite one.  

In some cases, programs need to contain some non-logical fragments, for
instance for input-output.  But the practice shows that Prolog makes possible 
building programs which are to a substantial extent declarative; in other 
words,
a substantial part of such program is a logic program. 
Numerous examples are given in the textbooks, for instance 
\cite{Sterling-Shapiro-shorter}.
For a more formal discussion of this issue see
\cite{drabent.tplp18}.

It should be noted that the operational, low-level approach to Prolog programming 
is often overused.  
In such programs it is not the declarative semantics that matters.
A typical example is the red cut \cite{Sterling-Shapiro-shorter}
  -- a programming technique which 
is based on pruning the search space; 
the program has undesired logical consequences, which are however not
computed due to the pruning.
Understanding such program substantially depends
on its 
operational semantics.  And understanding the operational semantics is 
usually more
difficult than that of the declarative semantics. 
In particular, examples of programs with the red cut are known,
for which
certain choices of the initial query lead to unexpected results
\cite[p.\,202-203]{Sterling-Shapiro-shorter}, 
\cite[Chapter 4]{Clocksin.Mellish.ed5}.
In seems that some Prolog textbooks over-use such style of programming
(like \cite{Clocksin.Mellish.ed5,Bratko.book.ed4}, at least in their earlier
editions).

%
%

%
%

\enlargethispage{.75ex}
\vspace{-.6ex} 

\paragraph
{Debugging Tools of Prolog.}
We begin with a terminological comment.  Often the term ``debugging'' is
related to locating errors in programs.  However its meaning is wider;
it also includes correcting errors.  So a better term for locating errors is
{\em diagnosis}.  However this text still does not reject the first usage,
as it is quite common.

Despite Prolog has been designed mainly as an implementation of logic
programming, its debugging tools work solely in terms of the operational
semantics.
So all the advantages of declarative programming are lost when it comes to
locating errors in a program.
The Prolog debugger is basically a tracing tool.
It communicates with the programmer only in terms of the operational semantics.
She (the programmer) must abandon the convenient high abstraction
level of the declarative semantics and think about her program in operational
terms. 

\vspace{-.6ex} 
\paragraph{Declarative Diagnosis.}
In principle, it is well known how to locate errors in logic programs 
declaratively, i.e.\ abstracting from the operational semantics
(see e.g.\ \cite[Section 7]{drabent.tocl16} and the references therein). 
The approach is called {\em declarative diagnosis} (and was introduced under
a name {\em algorithmic debugging} by Shapiro \cite{shapiro.book}).
Two kinds of errors of the declarative semantics of a program are dealt with:
{\em incorrectness} -- producing results which are wrong according to the
specification, and {\em incompleteness} -- not producing results which are
required by the specification.
We learn about an error by encountering a {\em symptom} -- a wrong or 
missing answer obtained at program testing.
Given a symptom,
an incorrectness (respectively incompleteness) diagnosis algorithm
semi-automatically locates an error in the program,
asking the user some queries about the specification.

Unfortunately, declarative diagnosis was not adapted in practice.
No tools for it are included in current Prolog systems.

\paragraph{Intended Model Problem.}
A possibly main reason for lack of acceptance of declarative diagnosis was
discussed in \cite[Section 7]{drabent.tocl16}.  
Namely, declarative diagnosis requires that the programmer exactly knows the
relations to be defined by the program.
Formally this means that the programmer knows
the least Herbrand model of the intended program.
(In other words, the least Herbrand model is the specification.)
This requirement turns out to be unrealistic.  For instance, in an insertion
sort program we do not know how inserting an element into an unsorted list
should be performed.  This can be done in any way, as the algorithm inserts
elements only into sorted lists.
Moreover, this can be done differently in various versions of the program.
See \cite{drabent.tplp18} for a more realistic example.%
\footnote{%
In the main example of  \cite{drabent.tplp18}, the semantics of a particular
predicate differs at various steps of program development.
}
Let us call this difficulty {\em intended model problem}.

Usually the programmer knows the
intended least Herbrand model of her program only approximately.
She has an {\em approximate specification}\/: 
she knows a certain superset $S_{corr}$ and a certain subset $S_{\it c o m p l}$
 of the intended model.  
The superset tells what may be computed, and
the  subset -- what must be computed.
Let us call the former, $S_{corr}$,
the {\em specification for correctness} and the latter, $S_{\it c o m p l}$,
the {\em specification for completeness}.  Thus the program should be correct
with respect to the former specification and complete
with respect to the latter:  
$S_{\it c o m p l}\subseteq M_P\subseteq S_{corr}$ (where $M_P$ is the least
Herbrand model of the program).
In our example, 
it is irrelevant how an element is inserted into an unsorted list; thus the
specification for correctness would include all such possible insertions
(and the specification for completeness would include none).

Now it is obvious that when diagnosing incorrectness the programmer should
use the specification for correctness instead of the intended model,
and the specification for completeness should be used when diagnosing
incompleteness \cite{drabent.tocl16}. 
The author believes that this approach can make declarative diagnosis useful
in practice.

Intended model problem was possibly first noticed by
Pereira \cite{Pereira86-short}.  He introduced the notion of {\em inadmissible}
atomic queries.  A formal definition is not given.%
\footnote{%
   ``a goal is admissible if it complies with the intended use of the procedure
   for it -- i.e. it has the correct argument types -- irrespective of whether
   the goal succeeds or not'' (p.\,6 of the extended version of
   \cite{Pereira86-short}).
}
We may suppose that ground inadmissible atoms are those from 
$S_{corr}\setminus S_{\it c o m p l}$. 
Generally, this notion
is not declarative; an inadmissible atom seems 
to be one that should not appear as a selected atom in an LD-tree
of the program.

Naish \cite{DBLP:conf/acsc/Naish00} proposed a 3-valued diagnosis scheme.
The third value, {\em inadmissible}, is related to the search space of
a diagnosis algorithm, and to its queries.  The form of queries depends on
the particular algorithm, e.g.\ 
it may be an atom together with its computed answers.
So the third value is not (directly) related
to the declarative semantics of programs.  It turns out that applying the
scheme to incorrectness diagnosis (\cite[Section 5.1]{DBLP:conf/acsc/Naish00})
boils down to standard diagnosis w.r.t.\ $S_{corr}$, 
and applying it to incompleteness diagnosis 
(\cite[Section 5.2]{DBLP:conf/acsc/Naish00}) -- to the standard diagnosis
w.r.t.\ $S_{\it c o m p l}$ (where $S_{\it c o m p l}$ is the set of correct 
atoms, and
$S_{corr}$ is the set of correct or inadmissible ones).
So introducing the 3-valued scheme seems unnecessary (at least for
incorrectness and incompleteness diagnosis).

\paragraph{This Paper.} 
The role of this paper is to find if, how, and to which extent the Prolog
debugger can be used as a tool for declarative logic programming.
We focus on the debugger of SICStus Prolog.  We omit its advanced debugging
features, which are sophisticated, but seem not easy to learn and not known
by most of programmers.

The paper is organized as follows.
The next section deals with the Prolog debugger and the information it can
provide.  Section \ref{sec.diagnosis} discusses applying the debugger for
diagnosing incorrectness and incompleteness.  
Examples, missing here, are presented in
 \cite{drabent.lopstr19.examples}.
The last section contains
conclusions. 

\section{Prolog Debugger}

In this section we present the Prolog debugger and try to find out how to
use it to obtain the information necessary from the point of view of
declarative programming.
First we relate the computation model used by the debugger to the standard
operational semantics (LD-resolution).  We also formalize the information
needed for incorrectness and incompleteness diagnoses.  
For incorrectness diagnosis, given an atomic answer $A$ we need to know
which clause $H\gets\seq B$ have been used to obtain the answer $A$
($A$ is an instance of $H$), and which top-level atomic answers
(instances of $\seq B$) have been involved.
  For incompleteness diagnosis, the
related information is which answers have been computed for each selected 
instance of each body atom $B_i$ of each clause  $H\gets\seq B$ resolved with
a given atomic query $A$.
In Section \ref{sec.deb.output} we describe %
 the messages of the debugger.
Section \ref{sec.tracing} investigates how to extract from the debugger's
output the information of interest.

 \enlargethispage{.6ex}

\subsection{Byrd Box Model and LD-resolution}
\label{sec.Byrd}
The debugger refers to the operational semantics of Prolog in terms of a 
``Byrd box model''.   Roughly speaking, the model assigns four ports to each 
atom selected in LD-resolution.  From a programmer's point of view such atom
can be called a {procedure call}.  The model is usually easily understood by
programmers.  However it will be useful to relate it here to LD-resolution,
and to introduce some additional notions.
In this paper, we often skip \mbox{``LD-''} and by ``derivation'' we mean
``LD-derivation'' (unless stated otherwise).

\paragraph{Structuring LD-derivations.}
Let us consider a (finite or infinite) LD-derivation $D$ with queries
$Q_0,\linebreak[3]Q_1,\linebreak[3]Q_2\ldots$, the input clauses $C_1,C_2,\ldots$, and the mgu's
 $\theta_1,\theta_2,\ldots$.
By a {\bf procedure call}  of $D$ we mean the atom selected in a query of $D$.
Following \cite{DM88,drabent.faoc16},
we describe a fragment of $D$ which may be viewed as the evaluation of 
a given procedure call $A$.

\begin{definition}\rm
\label{def.subderivation}
    Consider a query $Q_{k-1} = A,\seq[m]B$ ($m\geq0$) in a derivation $D$ as above.
  If $D$ contains a query $Q_l = (\seq[m]B)\theta_k\cdots\theta_l$, $k\leq l$, 
  then the call $A$ (of  $Q_{k-1}$) {\bf succeeds} in $D$.

  In such case,
  by the {\bf subderivation} for $A$ (of  $Q_{k-1}$ in $D$)
  we mean the fragment
  of $D$ consisting of the queries $Q_i$ where $k-1\leq i\leq l$,
  and for $k-1\leq i< l$ each $Q_i$ contains more than $m$ atoms.%
  \footnote{%
    Thus each such $Q_i$ is of the form 
    $\seq[m_i]A,(\seq[m]B)\theta_k\cdots\theta_i$ where\linebreak $m_i>0$.\linebreak[3]
    This implies that the least $l>k$ is taken such that $Q_l$ is of the
    form $(\seq[m]B)\theta_k\cdots\theta_l$.
  }
  We call such subderivation {\bf successful}.
  The (computed) {\bf answer} for $A$ (of  $Q_{k-1}$ in $D$) is 
  $A\theta_k\cdots\theta_l$.

  If  $A$ (of  $Q_{k-1}$) does not succeed in $D$ then the
  {\bf subderivation} for $A$ (of  $Q_{k-1}$ in $D$) is the fragment
  of $D$ consisting of the queries $Q_i$ where $k-1\leq i$.

  By a {\em subderivation} (respectively an {\em answer}) for $A$ of $Q$ in
  an LD-tree \TT 
  we mean a subderivation (answer) for $A$ of $Q$ in a branch $D$ of \TT.
\end{definition}

Now we structure a subderivation $D$ for an atom $A$ by distinguishing in $D$
top-level procedure calls.
Assume $A$ is resolved with a clause $H\gets \seq A$ in the first step of $D$.
If then an instance of $A_i$ becomes a procedure call, we call it a top-level
call.  More precisely:

\begin{definition}\rm
  \label{def.top-level}
 Consider a subderivation $D$ for $A$,
  with first two queries \linebreak
  {$Q_{k-1} = A,Q'$}
   and $Q_k = (\seq A,Q')\theta_k$, where $n>0$.  So
  $\seq A$ is the body of the clause used in the first step of the
  subderivation.
  Let $|Q_k|$ be the length of $Q_k$ (the number of atoms in $Q_k$).
{\sloppy\par}
  
  Consider an index $j$, $1\leq j\leq n$.
  If there exists in $D$ a query of the length $|Q_k|+1-j$ and
  $Q_{i_j} = (A_j,\ldots,A_n,Q')\theta_k\cdots\theta_{i_j}$ is the first such
  query then we say that 
  $A_j\theta_k\cdots\theta_{i_j}$ (of $Q_{i_j}$) is a {\bf top-level call} of
  $D$, 
  and the subderivation $D'$ for $A_j\theta_k\cdots\theta_{i_j}$ (of $Q_{i_j}$)
  in $D$ is a
  {\bf top-level subderivation} of $D$.  

\end{definition}
A top-level call of a subderivation $D$ for $A$
will be also called a top-level call {\rm for}~$A$.

Notice that if $A$ is resolved with a unary clause ($n=0$, and $D$ consists of
two queries) then $D$ has no top-level subderivations.
Also, if a top-level subderivation $D'$ of $D$ is successful 
then the last query of $D'$ is the first query
of the next subderivation, or it is the last query of $D$.

We are ready to describe what information to obtain from the debugger
in order to facilitate incorrectness and incompleteness diagnosis.
First we describe which top-level answers correspond to an answer for $A$;
we may say that they have been used to obtain the answer for $A$.

\begin{definition}\rm
  If subderivation $D$ for $A$ as in Def.\,\ref{def.top-level}
  is successful then it has $n$ top-level
  subderivations, for atoms  $A_j\theta_k\cdots\theta_{i_j}$ ($j=1,\ldots,n$).
  Their answers in $D$ are, respectively,
  $A_j' = A_j\theta_k\cdots\theta_{i_{j+1}}$
  (where $i_{n+1}$ is the index of the last query 
 $Q_{i_{n+1}} = Q'\theta_k\cdots\theta_{i_{n+1}}$ of $D$).
  In such case, by the {\bf top-level success trace} for $A$ (in $D$) we
  mean the sequence $\seq{A'}$ of the answers.
\end{definition}

Top-level success traces will be employed in incorrectness diagnosis.
For diagnosing incompleteness, we need to collect all the answers for each
top-level call.

\begin{definition}\rm
  Consider an LD-tree  \TT  with a node $Q$.  Let $A$ be the first
  atom of $Q$.  
  By the {\em top-level search trace} (or simply {\bf top-level trace})
  for $A$ (of $Q$ in \TT) we mean the set of pairs
  \[
  \left\{
  (B,\{\seq[k]B\}) \:\left|\:
  \begin{tabular}{l}
      $B$ is the first atom of a node $Q'$ of $\TT$, \\    
    $Q'$ occurs in a subderivation $D'$ for $A$ of $Q$ in \TT, \\
  $B$ is a top-level call of $D'$, \\
    $\seq[k]B$ are the answers for $B$ of $Q'$ in $\TT$  
  \end{tabular}
  \right.\right\}.
  \]

\end{definition}

\subsection{Debugger Output}
\label{sec.deb.output}
   For the purposes of this paper, this section should provide a sufficient
   description of 
   the debugger.  We focus on the debugger of SICStus.
   For an introduction and further information about the Prolog debugger
   see e.g.\ the textbook \cite{Clocksin.Mellish.ed5} or the manual
   {\tt http://sicstus.sics.se/}.

Prolog computation can be seen as traversal of an LD-tree.
The Prolog debugger reports the current state of the traversal by
displaying one-line items; such an item contains a single atom augmented by
other information.  A procedure call $A$ is reported as an item
\[
\Call n d {A}
\]
and a corresponding answer $A'=A\theta_k\cdots\theta_i$ as 
\[
\Exit n d {A'}
\]
Here $n, d$ are, respectively, the unique invocation number and the current
depth of the invocation;  we skip the details.  What is
important is that, given an Exit item, the invocation number uniquely
determines the corresponding Call item.

Note that a node in an LD-tree may be visited many times, and 
usually more than one item correspond to a single visit.
For instance,
to the last node $Q_l$ of a successful subderivation for $A$ (say that from
Definition \ref{def.subderivation}) there correspond, at least,
an Exit item with atom $A\theta_k\cdots\theta_l$ and
a Call item with atom $B_1\theta_k\cdots\theta_l$ (provided $m>0$).
Note that such a node is often the last query of more than one successful
subderivations (cf.\,Def.\,\ref{def.top-level}).
In such case other Exit items correspond to $Q_l$.  They are displayed in the
order 
which may be described as leaving nested procedure calls.  More formally,
the order of displaying the Exit items is that of the increasing lengths of
the corresponding successful subderivations.
(The displayed invocation depths of these items are decreasing consecutive
natural numbers.)

An Exit item is preceded by \mbox{\tt?} when backtrack-points exist between
the corresponding Call and the given Exit.  Thus more answers are possible
for (the atom of) this Call.

At backtracking the debugger displays Redo items of the form
\[
\Redo n d {A'}
\]
Such item corresponds to an Exit item with the same numbers $n,d$ and atom $A'$.
Both items correspond to the same node of the LD-tree.
The Redo item appears, speaking informally, when the answer $A'$ is abandoned,
and the computation of a new answer for the same query begins.
SICStus usually does not display a Redo item when the corresponding Exit item
was not preceded by {\tt?}.

A Fail item
\[
\Fail n d {A}
\]
is displayed when no (further) answer is obtained for $A$.
This
means that a node with $A$ selected is being left (and will not be visited
anymore).  The numbers and the atom in a Fail item are the same as those in
the corresponding Call item.
Both the Call and Fail items correspond to the same node of the LD-tree.

We described the output of the debugger of SICStus.
Commands of the debugger will be described when necessary.
The debuggers of most Prolog systems are similar.
However important differences happen.  For instance the debugger of SWI-Prolog 
({\small\tt http://swi-prolog.org/}) does not display the invocation numbers.
This may make difficult e.g.\ finding the Call item corresponding to a given
Exit item.
On the other hand, the debuggers or Ciao ({\small\tt http://ciao-lang.org/})
and Yap  ({\small\tt https://www.dcc.fc.up.pt/~vsc/yap/})
seem to display such numbers.
{\sloppy\par}

\subsection{Obtaining Top-level Traces}
\label{sec.tracing}
We are ready to describe how to obtain top-level traces using the Prolog
debugger. We first deal with the search trace.

\begin{algorithm}[All answers]\rm
  \label{alg.all.answers}
Assume that we are at a Call port; the debugger displays
  \[
  \Call n d B  %
  \]
We show how to obtain all the answers for $B$.  Do repetitively the following.
\begin{enumerate}
\item 
Type \ss to skip the details of processing the query $B$ and to go to the
corresponding  {\tt Exit} or {\tt Fail} port.

\item
  If the obtained port is \ {\Exit n {\hspace{-.5em}d} {B'}} \ then $B'$ is a
  computed answer for $B$. 
Type \jj\rr (to jump to the {\tt Redo} port; { \Redo n {\hspace{-.5em}d} {B'}}
\ is displayed).  Repeat (step 1) to compute further answers

If the obtained port is a {\tt Fail} then all the answers have been obtained.
To come back to the initial {\tt Call} port, type \rr.
\end{enumerate}

\end{algorithm}

An alternative to using this algorithm is to simply run Prolog on query $B$
(e.g.\ using the ``break'' option of the debugger).

\renewcommand*{\myvspace}{\vspace{-.7ex}}
\begin{algorithm}[Top-level trace]\rm
  \label{alg.search.trace}
  Assume that we are at a Call port
  \[
  \Call n d A  %
  \]
  We show a way of obtaining the top-level search trace for $A$.
  Repetitively do the following.
  \begin{enumerate}
    \label{trace1.step1}
    \item
      If an item
\myvspace\myvspace\myvspace
      \[
      \Call {n} {d} {A} 
     \qquad\quad \mbox{ or } \qquad\quad      \Redo {n} {d} {A'}
      \]
      is displayed then
    type \enter to make one step of computation.%
\footnote{%
In the case of {\tt Call}
 there are three possibilities.
    If the result is an item    \    \Call {n_1\hspace{-.6em}} {d{+}1} {B_1} \
    then $B_1$ is an instance of the first atom of the body of the clause
    used in the resolution step.
    Obtaining 
    \ \Exit {n\hspace{-.7em}} d {A'} \
    means that a unary clause was used 
    and $A$ succeeded immediately.
    Obtaining \ \Fail {n\hspace{-.7em}} d {A} \ means that $A$ failed
    immediately, as it was not unifiable with any clause head.

    In the case of  {\tt Redo:}\,$A'$, we deal with backtracking after
    having obtained an answer $A'$ for $A$.  Then
    there is a fourth possibility:
    obtaining a {\tt Redo:}\,$B_j'$ item, where $B_j'$ is an answer obtained
    for (an instance of) an atom $B_j$ from the body of the clause used to
    obtain the answer $A'$
}

    \item
      \label{trace.end}
      If 
      \[ 
      \Fail n d {A}
      \]  
      is displayed then the search is completed.  The trace has been obtained.

    \item
      If
      \[ 
      \Exit n d {A'}
      \]  
      is displayed then type \jj\rr (to jump to the {\tt Redo} port of $A$,
      in order to continue the search).

    \item
      \label{step.callBi}
      If
\myvspace
      \[
      \Call {n_i} {d{+}1} {B_i} 
      \]
      is displayed then
      employ Algorithm \ref{alg.all.answers} to obtain the answers for
      $B_i$.  Query $B_i$ together with the answers is an element of the
      top-level trace for $A$.
      
\smallskip
      Now we are again at the same {\tt Call:}$B_i$ item.  Type \ss
      to arrive at the first answer for $B_i$, (or to a {\tt Fail} if there
      is none).

    \item
      If
      \[
      \Exit {n_i} {d{+}1} {B_i'} \qquad\quad \mbox{ or } \qquad\quad
      \Fail {n_i} {d{+}1} {B_i}
      \]
      is displayed then
      type \enter, to make a single step.%
\footnote{%
 After an {\tt Exit}, this leads to a {\tt Call:}$B_{i+1}$ item, or to an 
 {\tt Exit:}$A'$ item;  the latter when $B_i$ is (an instance of) the last
 atom of the used clause.  
 After a {\tt Fail}, this leads (in a simple case) to a  {\tt Redo:}$B_{i-1}$.

 Here $B_{i-1},B_{i},B_{i+1}$ are instances of three consecutive atoms of the
 used clause.  }

    \item
      \label{step.redoBi}
      If
      \[
      \Redo {n_i} {d{+}1} {B_i'}
      \]
      is displayed then
      type \ss
      to arrive at the next answer for $B_i$ (or to a {\tt Fail} if there
      is none).

  \end{enumerate}
\end{algorithm}

The algorithm outputs the same answers ({\tt Exit} items) twice (by Algorithm
\ref{alg.all.answers} and after an \ss at steps \ref{step.callBi} and
\ref{step.redoBi}).  So all the details of the trace are displayed even if 
we do not invoke Algorithm \ref{alg.all.answers}.  But obtaining the
top-level trace from such output seems too tedious; we
need to group each query with its answers
(e.g. by sorting by the invocation numbers),
and remove
unnecessary items.  This can be done by a shell command
{\tt\,cut -b 2- | \linebreak[3] sort -nk 1 \linebreak[3] | 
      egrep 'Call:|Exit:'\,}.

\begin{algorithm}[Top-level success trace]\rm
  \label{alg.success.trace}
  Assume that we obtained an Exit item containing an answer $A'$.
  The item corresponds to the last query of a successful subderivation $D$ for an
  atom $A$.  In order to extract from the debugger output the top-level
  success trace
  for $A$ in $D$, we need that the debugger has displayed the Call and Exit items
  containing the top-level calls of $D$ and the corresponding answers.
  If this is not the case then, at the \ \Exit{n\hspace{-.5em}} d {A'} \ 
  item, type \rr to arrive to the corresponding Call item,
 \ \Call{n\hspace{-.5em}} d {A}. \ 
  Then perform 
  Algorithm \ref{alg.search.trace} until arriving again to the {\tt Exit:}\,$A'$ item
  (all the invocations of Algorithm \ref{alg.all.answers} may be skipped).

  To select a top-level success trace from the printed debugger items, do
  repetitively the following.  The trace will be constructed backwards.
  Initially the current item is \ \Exit{n\hspace{-.5em}} d {A'}.
  Repetitively do the following:

  \begin{quote}
\addtolength{\leftskip}{-1em} %
    The current item is
    \[
    \Exit {n} {d} {A'} \qquad\quad \mbox{ or } \qquad\quad
    \Call {n_j} {d{+}1} {B_j}
    \]
    Consider the preceding item.
    If the immediately preceding item is
    \[
    \Exit {n_{j'}} {d{+}1} {B_{j'}'}
    \]
    then $B_{j'}'$ is obtained as an element of the success trace.  Find the
    corresponding 
    \[
    \Call {n_{j'}} {d{+}1} {B_{j'}}
    \]
    item, and make it the current item 

    Otherwise, the preceding item is
    \[
    \Call {n} {d} {A} \qquad\quad \mbox{ or } \qquad\quad
    \Redo {n} {d} {A''}
    \]
    and all the elements of the top-level success trace for $A$ have been found.
  \end{quote}
\end{algorithm}

The construction of a top-level success trace can be made more efficient,
 by re-starting the computation with $A'$ as
  the initial query.  Then the search space to obtain a success of $A'$ (and
  the corresponding top-level success trace) may be substantially smaller than that
  for original atomic query from the Call item.

\section{Diagnosis}
\label{sec.diagnosis}
This section first discusses diagnosis of incorrectness, and then that of
incompleteness.  In each case we first present the diagnosis itself, and then
discuss how it may be performed employing the Prolog debugger.

\subsection{Diagnosing Incorrectness}
\label{sec.incorrectness.diagnosis}
A {\em symptom} of incorrectness is an incorrect answer of the program.
More formally, consider a program $P$ and an Herbrand interpretation
$S_{corr}$, which is our specification for correctness.
A symptom is an answer
$Q$ such that  \mbox{$S_{corr}\notmodels Q$},
where $S_{corr}$ is the specification for correctness.
(In other words, $Q$ has a ground instance $Q\theta$ such that 
$Q\not\in S_{corr}$.)
When testing finds such a symptom, the role of diagnosis is to find the
error, this means the reason of incorrectness.  An error is a clause
of the program which out of correct (w.r.t.\  $S_{corr}$) premises produces
an incorrect conclusion.  More precisely:

\begin{definition}\rm
  \label{def.incorrectness.error}
  Given a definite program $P$  and a specification $S_{corr}$ (for correctness),
  an {\bf incorrectness error} is an instance
  \[
  H\gets \seq B \qquad (n\geq0)
  \]
  of a clause of $P$ such that $S_{corr}\models B_i$ for all
  $i=1,\ldots,n$, but  $S_{corr}\notmodels H$.

  An {\em incorrect clause} is a clause $C$ having an instance $C\theta$ which
  is an incorrectness error. 
\end{definition}
In other words, $C$ is an incorrect clause iff $S_{corr}\notmodels C$.
In what follows, by a {\em correct atom} we consider an atom $A$ such that 
$S_{corr}\models A$ (where $S_{corr}$ is the considered specification for
correctness). 

Note that we cannot formally establish which part of the clause is erroneous.
Easy examples can be constructed showing that an incorrect clause $C$ can be
corrected in various ways; 
and each atom of $C$ remains unchanged in some corrected version of $C$
\cite[Section 7.1]{drabent.tocl16}.  
\commenta{
  Possibility?  define something better based on maximal unchanged fragments of
  the clause. 
}

The incorrectness diagnosis algorithm is based on the notion of a proof tree,
called also implication tree.
\begin{definition}\rm
  Let $P$ be a definite program and $Q$ an atomic query.  A {\bf proof tree}
  for $P$ and $Q$ is a finite tree in which the nodes are atoms, the root is
  $Q$ and 

  \begin{center}
    if \quad
    \parbox{.27\textwidth}
           {\raggedright
             $\seq B$ are  the children of a node $B$
           }
           \quad then \quad
           \parbox{.35\textwidth}
                  {
                    $B\gets\seq B$ is an \\ instance of a clause of $P$
                  }
                  ($n\geq0$).
  \end{center}
\end{definition}
Note that
the leaves of a proof tree are instances of unary clauses of $P$.

Now diagnosing incorrectness is rather obvious.  If an atom $Q$ is 
a symptom
then there exists a proof tree for $P$, $Q$.  The tree must
contain an incorrectness error (otherwise the root of the tree is correct, i.e.\
$S_{corr}\models Q$).
A natural way of searching for the error, in other words an incorrectness
diagnosis algorithm, is as follows:
Begin from the root and,
recursively, check the children $\seq B$ of the current node whether they are
correct (formally, whether $S_{corr}\models B_i$).  If all of them are
correct, the error is found;
it is $B\gets\seq B$ (where $B$ the parent of $\seq B$).
Otherwise take an incorrect child $B_i$, and
continue the search taking $B_i$ as the current node.

Obviously, such search locates a single error.  So correcting the error does
not guarantee correctness of the program.%
\footnote{%
  This does not even guarantee
  that the symptom we began with would disappear -- there may be some other
  errors involved.
}
\subsection{Prolog Debugger and Incorrectness}
\label{sec.debugger.incorrectness}
Now we try to find out to which extent the algorithm described above can be
mimicked by the standard Prolog debugger.
Unfortunately, the debugger does not provide a way to construct a proof tree
for a given answer.  We can however employ top-level success traces to perform
a search similar to that done by the incorrectness diagnosing algorithm
described in Section \ref{sec.incorrectness.diagnosis}.

\paragraph
{A Strategy for Incorrectness Errors.}
\label{strategy.corr}
Here we describe how to locate incorrectness errors using the Prolog
debugger. 

\commenta{
  !!!
  Discuss equivalence of the error search based on proof trees with that based
  on top-level success traces
}

\begin{algorithm}\rm\
\label{alg.incorr1}
Assume that while tracing the program we found out an incorrect answer $A'$
(for a query $A$).  So we are at an Exit item containing $A'$.  
Type \rr to arrive to the corresponding Call item
 \ \Call {n\hspace{-.5em}} {d} {A}. \
Do repetitively the following:

  \begin{enumerate}
  \item 
   Construct the top-level success trace $\seq[m]{B'}$ for the subderivation $D$
    (for an atom $A$, where $A'$ is the answer for $A$ in $D$),
    as described in Algorithm \ref{alg.success.trace}.
  \item 
    Check whether the atoms of the trace are correct
    (formally, whether $S_{corr}\models B_i'$).  If all of them are, then
    the search ends.

    Otherwise take an item \ \Exit {n_i\hspace{-.5em}} {d{+}1} {B_i'}, \
    in which $B_i'$ is incorrect, 
    and find  the corresponding Call item
     \ \Call {n_i\hspace{-.5em}} {d{+}1} {B}. \
    Now repeat the search, with $A,A'$ replaced by, respectively, $B,B_i'$,
by typing a command  \jj\cc\,$n_i$, or 
    by starting new tracing from query $B$ 
    (in some cases \jj\cc\,$n_i$ does not lead to the expected Call item).
  {\sloppy\par}

  \end{enumerate}
The last obtained top-level success trace  $\seq[m]{B'}$ points out the
incorrect clause (Def.\,\ref{def.incorrectness.error})
 of the program.  The clause is  $C = H\gets\seq[m]B$ such that
the obtained answers are instances of the body atoms
of $C$: each $B_j'$ is an instance of $B_j$, for $j=1,\ldots,m$.
The head $H$ of $C$ is unifiable with the last call $B$ for which the
top-level success trace was built.
\end{algorithm}

Obviously, the algorithm can be improved by 
checking the correctness of each element $B_i'$ of the trace as soon as it is
located.  (So the success trace needs to be constructed only until an
incorrect element is found.)

The approach of Algorithm \ref{alg.incorr1} is rather tedious.  A more natural
way to locate incorrectness errors is as follows.

\begin{algorithm}\rm\
\label{alg.incorr2}
\vspace{-.5ex}
\begin{enumerate}
\item
  Assume, as above, that an incorrect answer $A'$ was found.
  Begin as in Algorithm \ref{alg.incorr1}:
  arrive to the {\tt Call:}\,$A$ that resulted in the incorrect
  answer, and start constructing a top-level search trace.

\item
  For each obtained item \ \Exit {n_i\hspace{-.5em}} {d{+}1} {B'} \  check if
  $B'$ is correct.  
\item
  If $B'$ is an incorrect answer, then restart the search from $B'$.
\item
  If no incorrect answer has appeared until arriving to the incorrect answer 
  $A'$ then the error is found.  It is the last clause $C$ whose head was
  unified with $A$ in the computation.  (Formally, an instance of $C$ is an
  incorrectness error.)

  The clause may be identified, as previously, by extracting the top-level
  success trace (for the subderivation that produced $A'$).
\end{enumerate}
\end{algorithm}

\paragraph{Comments.}
In Algorithms \ref{alg.incorr1},\,\ref{alg.incorr2},
it is often not necessary to know the (whole) top-level success trace to
identify the 
incorrect clause in the program.  In many cases, knowing the
last one or two answers of the trace is sufficient.  For instance, let
 \ \Call {n'\hspace{-.5em}} {d'} {B} \  be the last 
call for which top-level trace was inspected.
The last item displayed by the debugger is
\ \Exit {n'\hspace{-.5em}} {d'} {B'} \ 
(where $B'$ is incorrect).
Assume that the previous item is
\ \Exit {n_j\hspace{-.5em}} {d'{+}1} {B_j'}.
Then the top-level trace of interest is not empty, $B_j'$ is its last atom
and is an instance of the last body atom of an erroneous clause.  If the
program has only one such clause, then finding the rest of the top-level
success trace is unnecessary.

\smallskip

The error located by the second approach (Algorithm \ref{alg.incorr2}) may
be not the one that caused the 
initial incorrect answer $A'$.
This is because the search may go into a branch of the LD-tree distinct from
the branch in which $A'$ is produced.
Anyway, an actual error has been discovered in the program.  This outcome is
useful, as each error in the program should be corrected.

Note that the approach is complete, in the sense that the error(s) responsible
for $A'$ can be found.  This is due to the nondeterministic search performed by
the algorithm.  The error(s) will be located under some choice of incorrect
answers in the top-level search traces.

\smallskip
The search may be made more efficient if,
instead of tracing the original computation, we re-start it with an incorrect
answer as a query.
The corresponding modification (of both algorithms) is as follows.
Whenever an incorrect answer $B'$
is identified, instead of continuing the search for the corresponding call $B$,
one interrupts the debugger session and begins a new one by starting Prolog
with query $B'$.  The query will succeed with $B'$ (i.e.\ itself) as an answer,
but the size of the trace may be substantially smaller (and is never greater).
Moreover, any incorrect instance of $B'$ may be used instead of $B'$.

\smallskip
The Prolog debugger does not facilitate searching for the reason of
incorrectness.  Finding a top-level success trace is tedious and not
obvious.  In particular, there seems to be no way of skipping the
backtracking that precedes obtaining the wrong answer.
The abilities of the debugger make Algorithm \ref{alg.incorr2} preferable;
this approach in a more straightforward way uses what is offered by the
debugger. 

Looking for the reason of an incorrect answer is a basic task.
It is strange that such a task is not conveniently facilitated by the
available debugging tools.

\vspace{-.5ex}

\subsection{Diagnosing Incompleteness}

A specification for completeness is, as already stated, an Herbrand
interpretation which is the set of all required ground
answers of the program. 
A symptom of incompleteness is lack of some answers of the program.
More formally, given a program $P$ and a specification $S_{\it c o m p l}$,
by an incompleteness symptom we may consider a ground atom $A$ such that
$S_{\it c o m p l}\models A$ but $P\notmodels A$.  As a symptom is to be
obtained out of an actual computation, we additionally require that the
LD-tree for $A$ is finite.
We will consider a more general notion of a symptom:
%

%

%
%

\begin{definition}\rm
  Consider a definite program $P$ and a specification $S_{\it c o m p l}$ (for
  completeness).  Let $A$ be an atomic query for which an LD-tree is
  finite and let $\seq{A\theta}$ be the computed answers for $A$ from the tree.
  If there exists an instance $A\sigma\in S_{\it c o m p l}$ such that 
  $A\sigma$  is not an
  instance of any $A\theta_i$ ($i=1,\ldots,n$) then $A, \seq{A\theta}$
  is an {\bf incompleteness symptom} (for $P$ w.r.t.\ $S_{\it c o m p l}$).
\rm

\end{definition}
  We will often skip the sequence of answers, and say that $A$ alone is the
  symptom. 
  The definition can be generalized to non-atomic queries in an obvious way.

\begin{definition}\rm
  Let $P$ be a definite program, and $S_{\it c o m p l}$ a specification.
  A ground atom $A$ is {\bf covered} by a clause $C$ w.r.t.\ $S_{\it c o m p l}$
  if there exists a ground instance $A\gets\seq B$ of $C$ ($n\geq0$) 
  such that all the atoms  $\seq B$ are in $S_{\it c o m p l}$.

  $A$ is {\em covered by the program} $P$ (w.r.t.\ $S_{\it c o m p l}$) if $A$
  is covered by some clause $C\in P$.

\vspace{-4pt}  
\end{definition}
Informally, $A$ is covered by $P$ if it can be produced by a rule from $P$
out of some atoms from the specification.

If there exists an incompleteness symptom for $P$ w.r.t.\ $S_{\it c o m p l}$
then there exists an atom $p(\vec t)\in S_{\it c o m p l}$ 
uncovered by $P$ w.r.t.\ $S_{\it c o m p l}$
\cite{shapiro.book,drabent.tocl16}.  
Such an atom locates the error in $P$.  This is because
no rule of $P$ can produce $p(\vec t)$ out of atoms required to be produced.
This shows that the procedure $p$ (the set of clauses beginning with $p$)
is the reason of the incompleteness and 
has to be modified, to make the program complete.
Note that similarly to the incorrectness case, we cannot locate the error
more precisely.  Various clauses may be modified to make $p(\vec t)$ covered,
or a new clause may be added.  An extreme case is adding to $P$ a fact 
$p(\vec t)$. 

Incompleteness diagnosis means looking for an uncovered atom,
or -- more generally -- for an atom with an instance which is uncovered:
Such atom localizes the procedure of the program which is responsible for
incompleteness.

\begin{definition}\rm
Let $P$ be a definite program, and $S_{\it c o m p l}$ a specification.
An  {\bf incompleteness error} (for $P$ w.r.t.\ $S_{\it c o m p l}$) is an
atom that has an instance which is not covered
 (by $P$ w.r.t.\ $S_{\it c o m p l}$).
\end{definition}

Name ``incompleteness error''
may seem unnatural, but we find it convenient.

A class of incompleteness diagnosis algorithms employs the following idea.
Start with an atomic query $A$ (which is a symptom) and construct a top level
trace for it.  Inspect the trace, whether it contains a symptom $B$.  If so
then invoke the search recursively with $B$.  Otherwise 
$A$ is an incompleteness error; we located in the program
the procedure that is responsible for the incompleteness.
Such approach (see e.g.\ \cite{Pereira86-short,DNM89}) 
is sometimes called {\em Pereira-style} incompleteness diagnosis 
\cite{DBLP:journals/ngc/Naish92}.

\renewcommand*{\myvspace}{\vspace{-.6ex}}

\subsection{Prolog Debugger and Incompleteness}
We show how 
Pereira-style diagnosis may be done using the Prolog debugger.

\enlargethispage{1.4ex}

\begin{algorithm}
  [Incompleteness diagnosis]
\label{alg.incompleteness}
\rm \
Begin with a symptom $A$. 
Obtain the top-level search trace for $A$  (Algorithm  \ref{alg.search.trace}).
In the trace, check if the atom $B$ from a Call item
together with the answers $\seq B$ from the corresponding Exit items is an
incompleteness symptom.  If yes, invoke the same search starting from $B$.
If the answer is no for all Call items of the trace,
the search is ended as we located $A$ as an incompleteness error.

\end{algorithm}

\enlargethispage{.5ex}
\renewcommand*{\myvspace}{\vspace{-.6ex}}
\myvspace
\myvspace

\paragraph{Comments.}
Standard comments about incompleteness diagnosis apply here. 
To decrease the search space, it is useful to start the diagnosis
from a ground instance $A\theta\notin S_{\it c o m p l}$ of the symptom $A$
(instead of $A$ itself).  The same for each symptom $B$ found during the
search -- re-start the computation and the diagnosis from an appropriate
instance of $B$.

Often an incorrectness error coincides with an incompleteness error -- a wrong
answer is produced instead of a correct one.  The programmer learns about
this when facing an incorrect answer $B_i$ (appearing in a top-level trace). 
A standard advice in such case
\cite{DNM89,DBLP:journals/ngc/Naish92} is to switch to incorrectness
diagnosis.  This is because incorrectness diagnosis is simpler, and it
locates an error down to a program clause (not to a whole procedure, as
incompleteness diagnosis does).  The gain of such switch
is less obvious in our case, since 
the effort needed for incorrectness diagnosis (Algorithm \ref{alg.incorr2})
may be not smaller than that for incompleteness (Algorithm
\ref{alg.incompleteness}).

\myvspace

\section{Conclusions}

\myvspace

Prolog makes declarative logic programming possible -- programs may be
written and reasoned about in terms of their declarative semantics,
to a substantial extent abstracting from the operational semantics.
This advantage is lost when it comes to locating errors in programs, as the
Prolog debugger works solely in terms of the operational semantics.
We may say that 
logic programming would not deserve to be called a declarative programming 
paradigm if debugging had to be based on the operational semantics.

This paper is an attempt to study if and how the Prolog debugger can be used 
for declarative programming.  It presents how the debugger can be used to
perform incorrectness and incompleteness diagnosis.%
\footnote{We
    may informally present the underlying idea of this paper in a different way:
  To understand what
 the Prolog debugger can tell us about the declarative semantics of the program,
  we need to be able to obtain the following information.
  \ 1.~For a given atomic answer $A$, 
  what are the top-level answers that have lead to $A$?\
  (This is formalized as top-level success trace.)
  \ 2.~For a given atomic query $Q$, and for each top-level atomic query $B$ in
  the computation for $Q$, what are all the answers for $B$\/?
}
See \cite{drabent.lopstr19.examples} for examples.
%
  The debugger used is
that of SICStus;
 the presented approach may be difficult to apply with the debugger
 of SWI-Prolog, as the latter does not display unique invocation numbers
 (needed in incorrectness diagnosis,
Algorithms \ref{alg.success.trace}, \ref{alg.incorr1}).

The results are rather disappointing.  Declarative diagnosis based on the
Prolog debugger is tedious and unnatural.  Rather obvious information (like
the proof tree leading to a given answer, or a top-level success trace) is
impossible or difficult to obtain.
Possibly, this drawback
is a substantial obstacle for employing
declarative logic programming in practice.

This drawback particularly concerns incorrectness diagnosis.
Additionally, debugging of incorrectness seems more important than that of
incompleteness.  This is because incompleteness is often caused by producing
incorrect answers instead of correct ones.  Also, incorrectness diagnosis is
more precise,
as it locates a smaller erroneous fragment of the program
 than incompleteness diagnosis does.
Hence the first
step towards making Prolog debugging declarative
is to implement a tool supporting incorrectness diagnosis.
Experiments show that it is sufficient to provide a tool for convenient
browsing of a proof tree
(which provides an abstraction of the part of computation responsible for the
considered incorrect answer).
The Introduction contains a discussion about how to avoid the ``intended model
problem'', which is possibly
 the main reason why declarative diagnosis
of logic programs was
abandoned.  The author believes that the proposed solution
\cite{drabent.tocl16} can make declarative diagnosis useful in practice.
What is missing are tools.

\vspace{-1ex}

%

%

\bibliographystyle{splncs04}
\bibliography{biblopstr}

\end{document}